\input{aipcheck}

\documentclass[
    ,final            
  ]
  {aipproc}

\layoutstyle{6x9}
\def\tev{\ifmmode \mathop{\rm TeV}\nolimits\else {\rm TeV}\fi}
\def\gev{\ifmmode \mathop{\rm GeV}\nolimits\else {\rm GeV}\fi}
\def\mev{\ifmmode \mathop{\rm MeV}\nolimits\else {\rm MeV}\fi}
\def\kev{\ifmmode \mathop{\rm keV}\nolimits\else {\rm keV}\fi}
\def\ev{\ifmmode \mathop{\rm eV}\nolimits\else {\rm eV}\fi}
\def\ryd{\ifmmode \mathop{\rm Ry}\nolimits\else {\rm Ry}\fi}
\def\angst{\ifmmode\mathop{\rm\AA}\nolimits\else {\rm \AA}\fi}

\def\pepe{\mathop{\rm P.P.}}

\def\real{\mathop{\rm Re}}
\def\imag{\mathop{\rm Im}}

\def\dd{{\rm d}}

\def\degreec{\ifmmode\mathop{^\circ \rm C}\nolimits\else{$^\circ{\rm C}\;$}\fi}
\def\degreek{\ifmmode\mathop{^\circ \rm K}\nolimits\else{$^\circ{\rm K}\;$}\fi}
\def\degreef{\ifmmode\mathop{^\circ \rm F}\nolimits\else{$^\circ{\rm F}\;$}\fi}

\def\chidof{\ifmmode\mathop\chi^2/{\rm d.o.f.}\else $\chi^2/{\rm d.o.f.}\null$\fi}

\def\msbar{\ifmmode\mathop{\overline{\rm MS}}\else$\overline{\rm MS}$\null\fi}

\def\cmass{\ifmmode\mathop{\rm c.m.}\nolimits\else {\sl c.m.}\fi}
\def\lab{\ifmmode{\rm lab}\else {\sl lab.}\fi}

\def\degrees{\ifmmode{^\circ\,}\else $^\circ$\fi}
\def\feet{\ifmmode{\hbox{'}\,}\else '\fi}
\def\inches{\ifmmode{\hbox{"}\,}\else "\fi}

\def\lsim{\mathop{\setbox0=\hbox{$\displaystyle 
\raise2.2pt\hbox{$\;<$}\kern-7.7pt\lower2.6pt\hbox{$\sim$}\;$}
\box0}}
\def\gsim{\mathop{\setbox0=\hbox{$\displaystyle 
\raise2.2pt\hbox{$\;>$}\kern-7.7pt\lower2.6pt\hbox{$\sim$}\;$}
\box0}}

\def\frac#1#2{{#1\over#2}}
\def\dfrac#1#2{{\displaystyle{#1\over#2}}}
\def\tfrac#1#2{{\textstyle{#1\over#2}}}
\def\ffrac#1#2{\leavevmode
   \kern.1em \raise .5ex \hbox{\the\scriptfont0 #1}%
   \kern-.1em $/$%
   \kern-.15em \lower .25ex \hbox{\the\scriptfont0 #2}%
}%

\def\be{\begin{equation}}
\def\ee{\end{equation}}
\def\bea{\begin{eqnarray}}
\def\eea{\end{eqnarray}}

\begin{document}

\title{Consistency checks of different $\pi\pi$ scattering data sets
using forward dispersion relations}


\author{Jos\'e R. Pel\'aez}{
  address={Dept. F\'isica Te\'orica II. Universidad Complutense, 28040-Madrid, Spain}
}

\begin{abstract}
We review our evaluation of forward dispersion relations for direct
fits to the different, and often conflicting, $\pi\pi$
scattering experimental analyses. 
We find that
some of the most commonly used data sets
do not satisfy these constraints by several standard deviations. 
We also provide a consistent
$\pi\pi$ amplitude by improving a global fit
to data with these dispersion relations.
\end{abstract}
\maketitle



\vspace{-.4cm}
\noindent
A precise knowledge of the $\pi\pi$ scattering amplitude
provides crucial tests
for Chiral Perturbation Theory (ChPT), 
as well as information on
light meson spectroscopy, pionic atom decays and CP violation in kaons.
However, there are several $\pi\pi$ scattering data sets in the literature,
in conflict among themselves, even within
the same experiment. The reason is that the data
are extracted from
other reactions and thus with large theoretical and
systematic uncertainties.
Here we review our recent works \cite{3,4,5} where we checked 
dispersion relations on the
different sets of data and provided 
simple parameterizations of $\pi\pi$ scattering amplitudes
consistent with such requirements.

\vspace{-.7cm}
\subsection{ Fits to different sets of data}
\vspace{-.3cm}

\noindent
We 
first consider {\it fits to data} \cite{5} for the S0, S2, P waves, 
below $s^{1/2}\lsim 1\,\gev$.
We parameterize the phase shifts, $\delta(s)$,
taking into account the analytic properties,
zeros and
poles of the amplitude, in a conformal expansion of $\cot\delta(s)$.

The P wave, up to $\simeq1\,$GeV, comes from the following 
 fit to the pion form factor \cite{6}
\begin{eqnarray}
&&\cot\delta_1(s)=\dfrac{s^{1/2}}{2k^3}
(M^2_\rho-s)\left\{B_0+B_1\dfrac{\sqrt{s}-\sqrt{s_0-s}}{\sqrt{s}+\sqrt{s_0-s}}
\right\};\quad s_0^{1/2}=1.05\;\gev.
\label{(2.1a)}\\
&&B_0=\,1.069\pm0.011,\quad B_1=0.13\pm0.05,\quad M_{\rho}=773.6\pm0.9
\nonumber
\end{eqnarray} 
($s_0$ is the point where inelasticity begins to be nonnegligible). The fit
is seen in Fig.1a.

For the S2 wave 
at low energies, we first 
fix the Adler zero 
at $z_2=M_\pi$ and fit only the low energy data, 
$s^{1/2}<1.0\,\gev$; later on we allow $z_2$ to vary. We 
have
\begin{eqnarray}
&&\cot\delta_0^{(2)}(s)=\dfrac{s^{1/2}}{2k}\,\dfrac{M_{\pi}^2}{s-2z_2^2}\,
\left\{B_0+B_1\dfrac{\sqrt{s}-\sqrt{s{_0}-s}}{\sqrt{s}+\sqrt{s{_0}-s}}\right\},
\quad s{_0}^{1/2}=1.05\;\gev,
\\
&&B_0=\,-80.4\pm2.8,\quad B_1=-73.6\pm12.6;
\end{eqnarray}

\hspace*{-.5cm}
\begin{minipage}{1.0\linewidth}
  \includegraphics[height=0.333\linewidth,angle=-90]{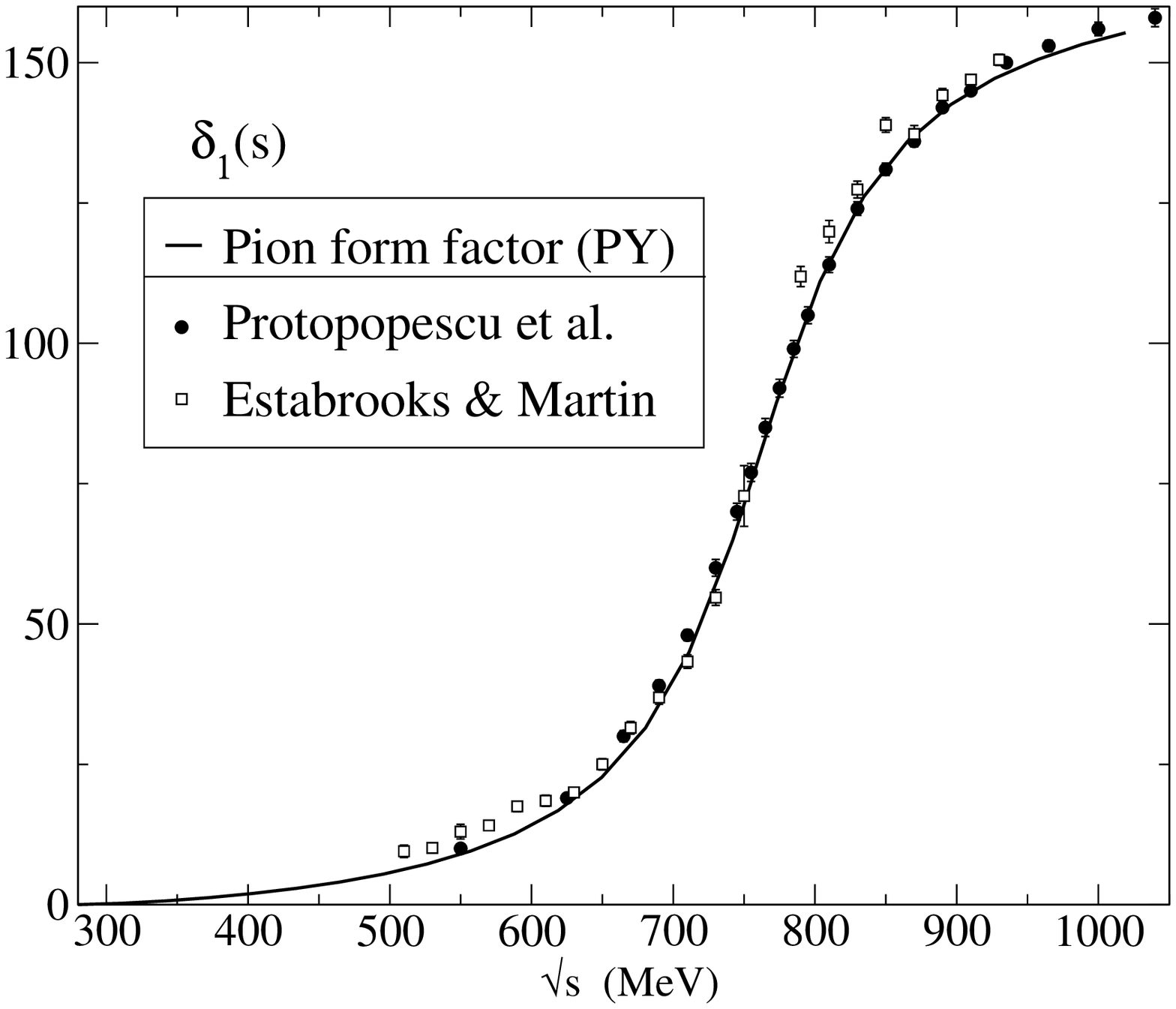}
  \hspace*{-.2cm}
  \includegraphics[height=0.333\linewidth,angle=-90]{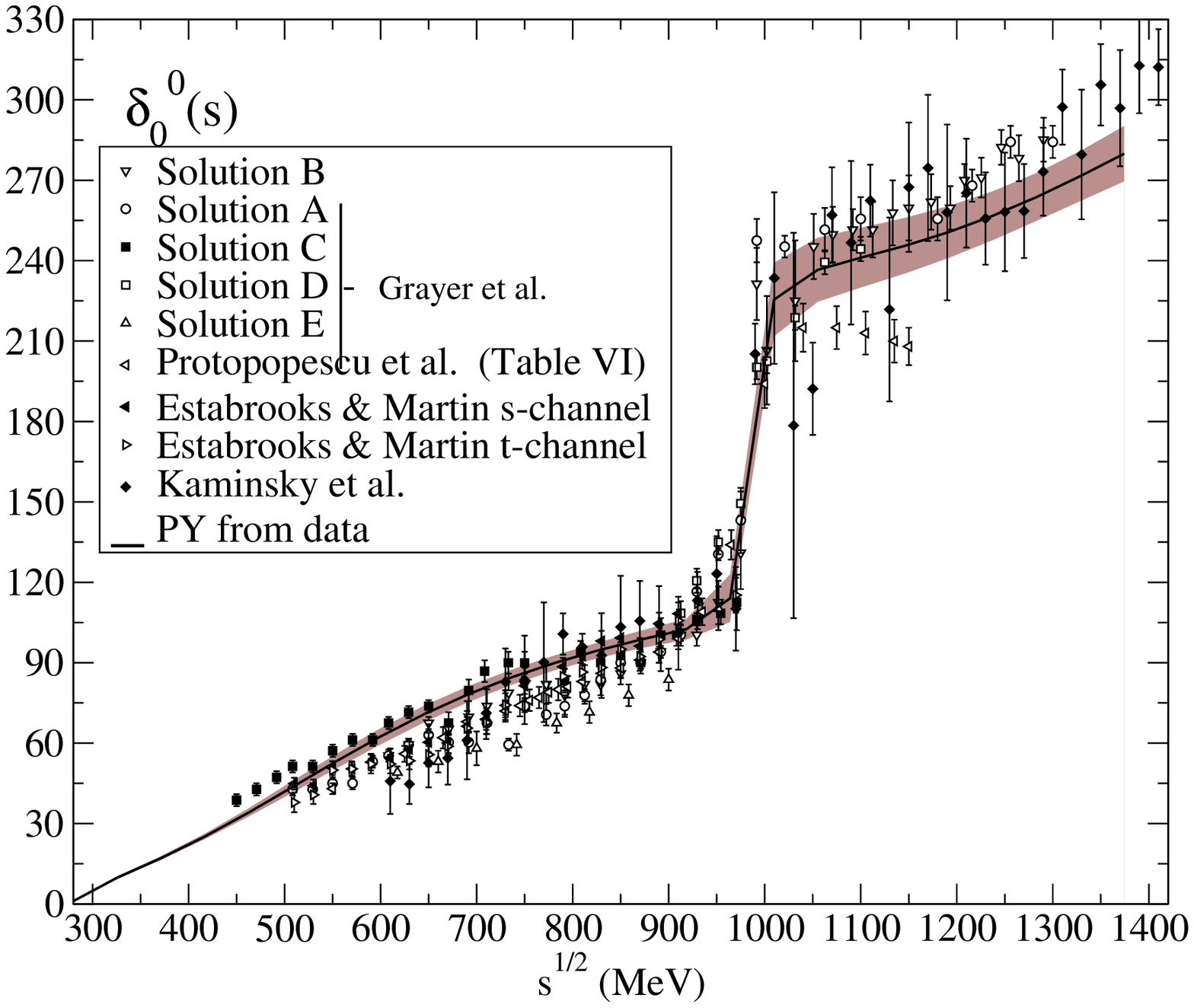}
  \hspace*{-.2cm}
  \includegraphics[height=0.333\linewidth,angle=-90]{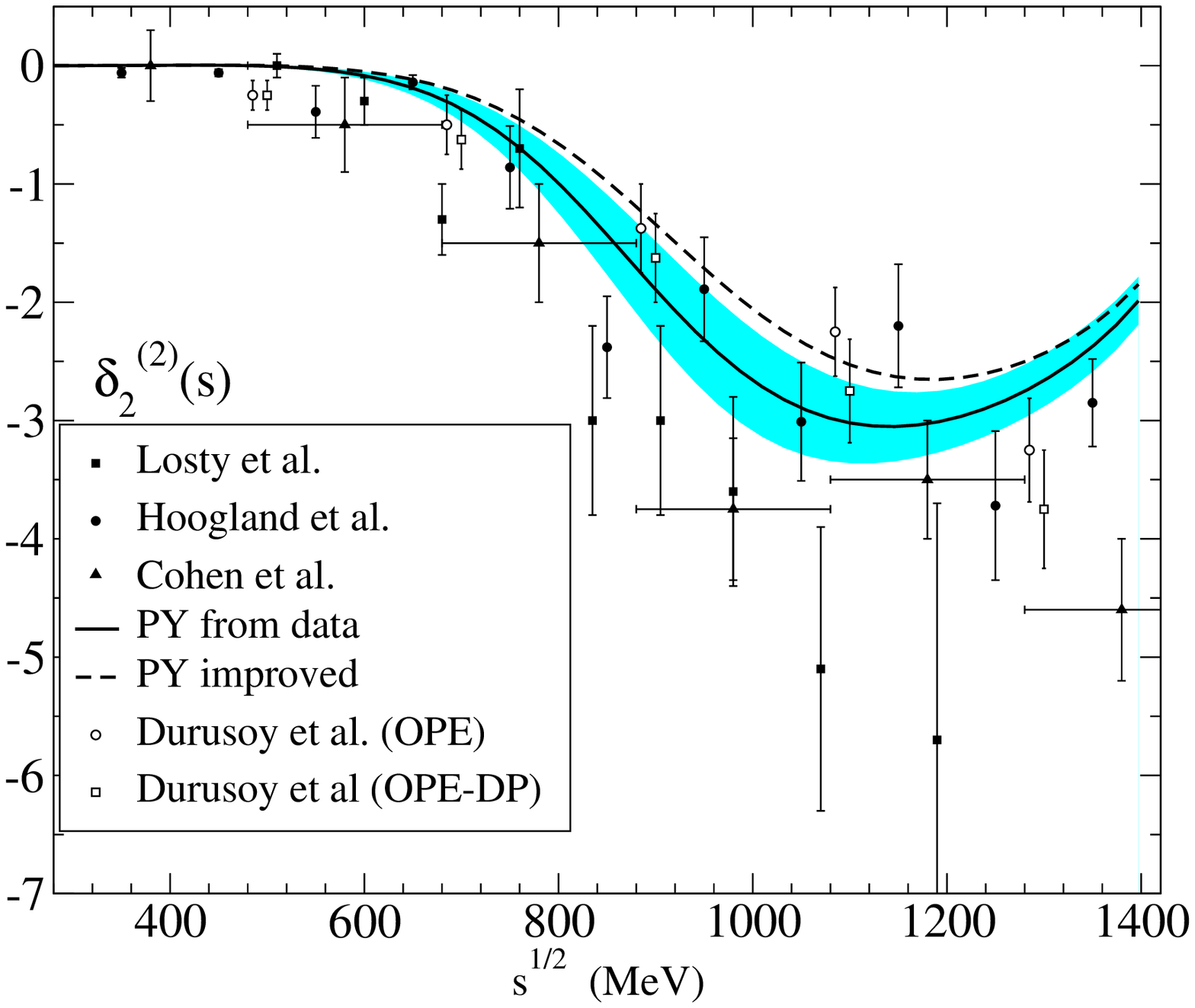}

{\footnotesize {\sc \bf FIGURE 1.} a) P wave $\pi\pi$ phase shifts \cite{9,10}, versus
    Eq.(\protect\ref{(2.1a)}) (solid line below 1~\gev), which is {\sl
      not} a fit to these data, but to the pion form factor \cite{6}.
    The uncertainty is the line thickness.  Above 1~\gev, the dotted
    line and error are as in \cite{5}. b) S0 phase shifts and error
    band as given by Eq.(\ref{(2.3a)}) below 1~\gev, and
    from \cite{5} above.  The $K_{l4}$ and $K_{2\pi}$ decay data are
    not shown. (see our \cite{5} for details). c) Continuous line: The
    fit to $I=2$, $D$-wave phase shift data.  Broken line: fit
    improved with dispersion relations.  The experimental points are
    from \cite{7}. }
\end{minipage}
\vspace{.2cm}

Except for the very reliable $K_{l4}$ and $K\to2\pi$ decay
experiments \cite{8}, that we always fit, 
the S0 data is very confusing and thus we have adopted two approaches.
\emph{In the first method}, 
the ``{\it global fit}'', we
fit averaged phase shift data between
 $0.81\,\gev\leq s^{1/2}\leq0.97\,\gev$,
(where the experiments agree within 
$1.5\,\sigma$), composing their errors carefully.
The fit, shown in Fig.1.b, is  valid 
for $s^{1/2}\leq0.95\,\gev$, and with 
the Adler zero fixed at $z_0=M_\pi$, corresponds to
\begin{eqnarray}
&&\cot\delta_0^{(0)}(s)=\,\dfrac{s^{1/2}}{2k}\,\dfrac{M_{\pi}^2}{s-\tfrac{1}{2}z_0^2}\,
\dfrac{M^2_\mu-s}{M^2_\mu}\,
\left\{B_0+B_1\dfrac{\sqrt{s}-\sqrt{s_0-s}}{\sqrt{s}+\sqrt{s_0-s}}\right\},
\label{(2.3a)} \\
&&B_0=y-x;\quad B_1=6.62-2.59 x;\quad y=21.04\pm0.70,\quad x=0\pm 2.6.
\nonumber
\end{eqnarray}

\emph{The second method} is to fit only  $K_{l4}$ and $K\to2\pi$ data, or 
to add to them the data from the various experimental analyses separately.
The results can be found in Table~1.

The D2 inelasticity is negligible
below $4M^2_\rho$.
A pole term is needed
since 
the data \cite{7} gives
a small negative phase above $\sim500\,\mev$, but
the Froissart--Gribov representation, 
yields \cite{12} a positive scattering length,
$a_2^{(2)}=(2.72\pm0.36)\times10^{-4}\,M_{\pi}^{-5}$,
which is included in the fit.
Also, the inflection seen in data around 1~\gev\ 
asks for a third order conformal expansion. So 
we write
\begin{eqnarray}
\cot\delta_2^{(2)}(s)=
\dfrac{s^{1/2}}{2k^5}\,\Big\{B_0+B_1 w(s)+B_2 w(s)^2\Big\}\,
\dfrac{{M_\pi}^4 s}{4({M_\pi}^2+\Delta^2)-s},\;
w(s)=\dfrac{\sqrt{s}-\sqrt{s_0-s}}{\sqrt{s}+\sqrt{s_0-s}}.\nonumber
\label{(2.4a)}
\end{eqnarray}
And we find
$B_0=(2.4\pm0.3)\times10^3$, $B_1=(7.8\pm0.8)\times10^3$,
 $B_2=(23.7\pm3.8)\times10^3$,
$\Delta=196\pm20\,\mev$. The fit is shown in Fig.1.c.

For brevity,  we do not discuss here the D0 and F waves 
as they do not present special features, nor the
intermediate energy region $1\,\gev\geq s^{1/2}\geq1.42\;\gev$,
but a detailed account of their parameterizations can be found in \cite{5}.

We also need the 
 imaginary part of the scattering amplitude at $s^{1/2}\geq1.42\;\gev$,
that we  take  
from a Regge fit to experimental data ( see \cite{4} for details
and the slightly improved rho residue of \cite{5}),
shown in Fig.2. As discussed in  \cite{4,5}, 
standard Regge factorization
describes experiment \cite{14,4} and 
is consistent with
crossing  sum rules.

\begin{minipage}{1.0\linewidth}
  \hspace*{-.7cm}
  \includegraphics[height=.28\textheight]{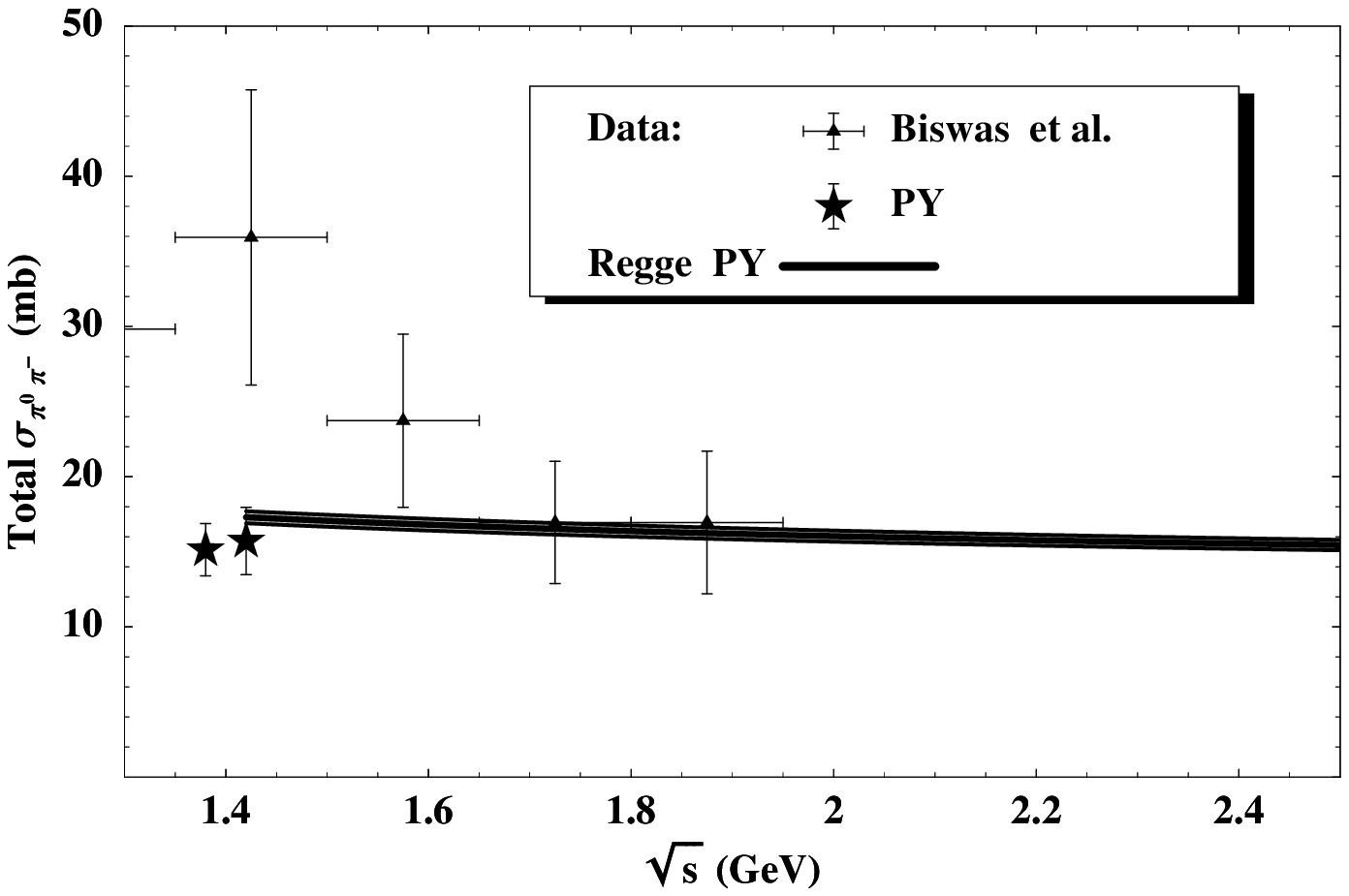}
  \includegraphics[height=.28\textheight]{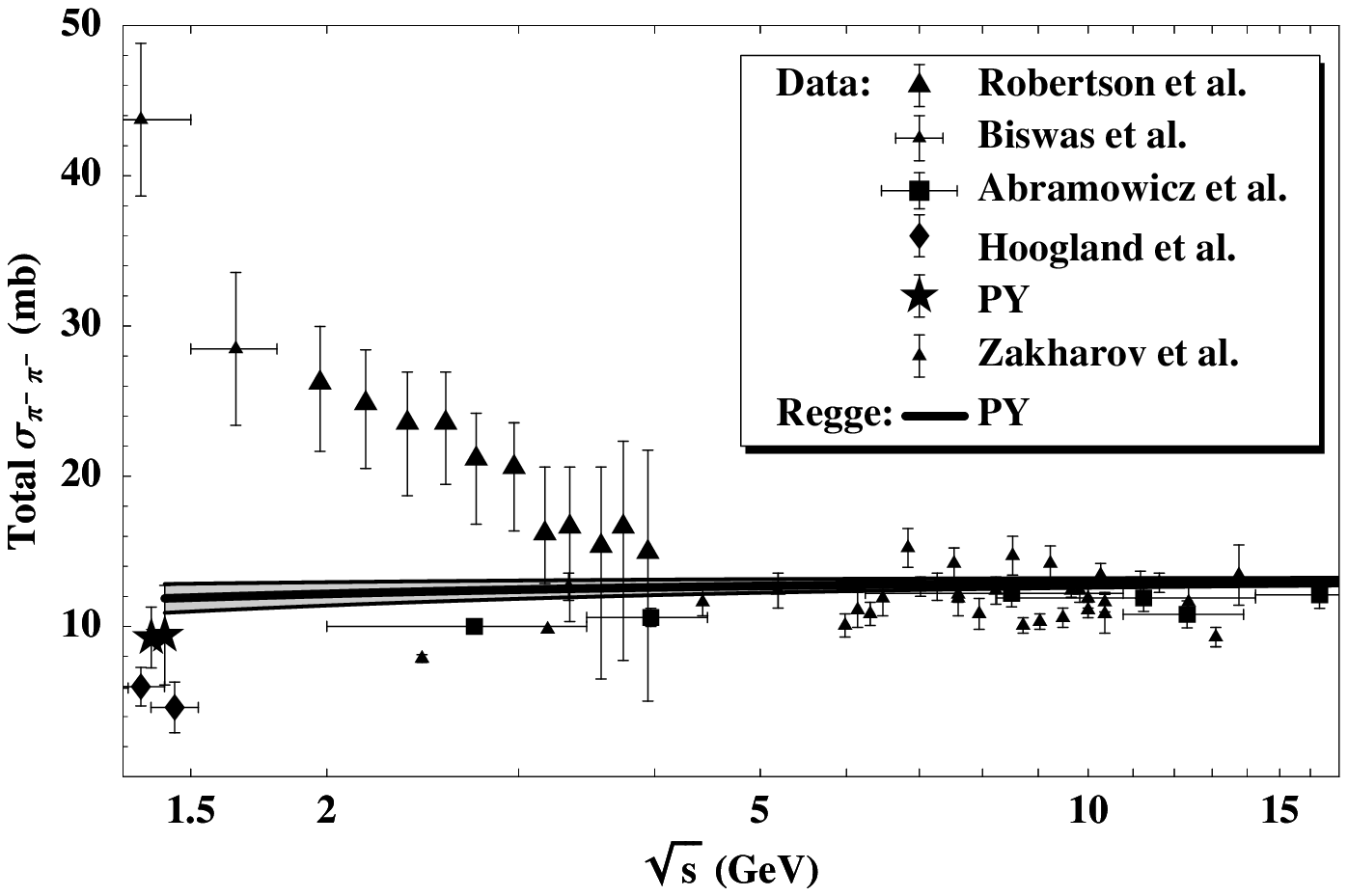}
  \includegraphics[height=.28\textheight]{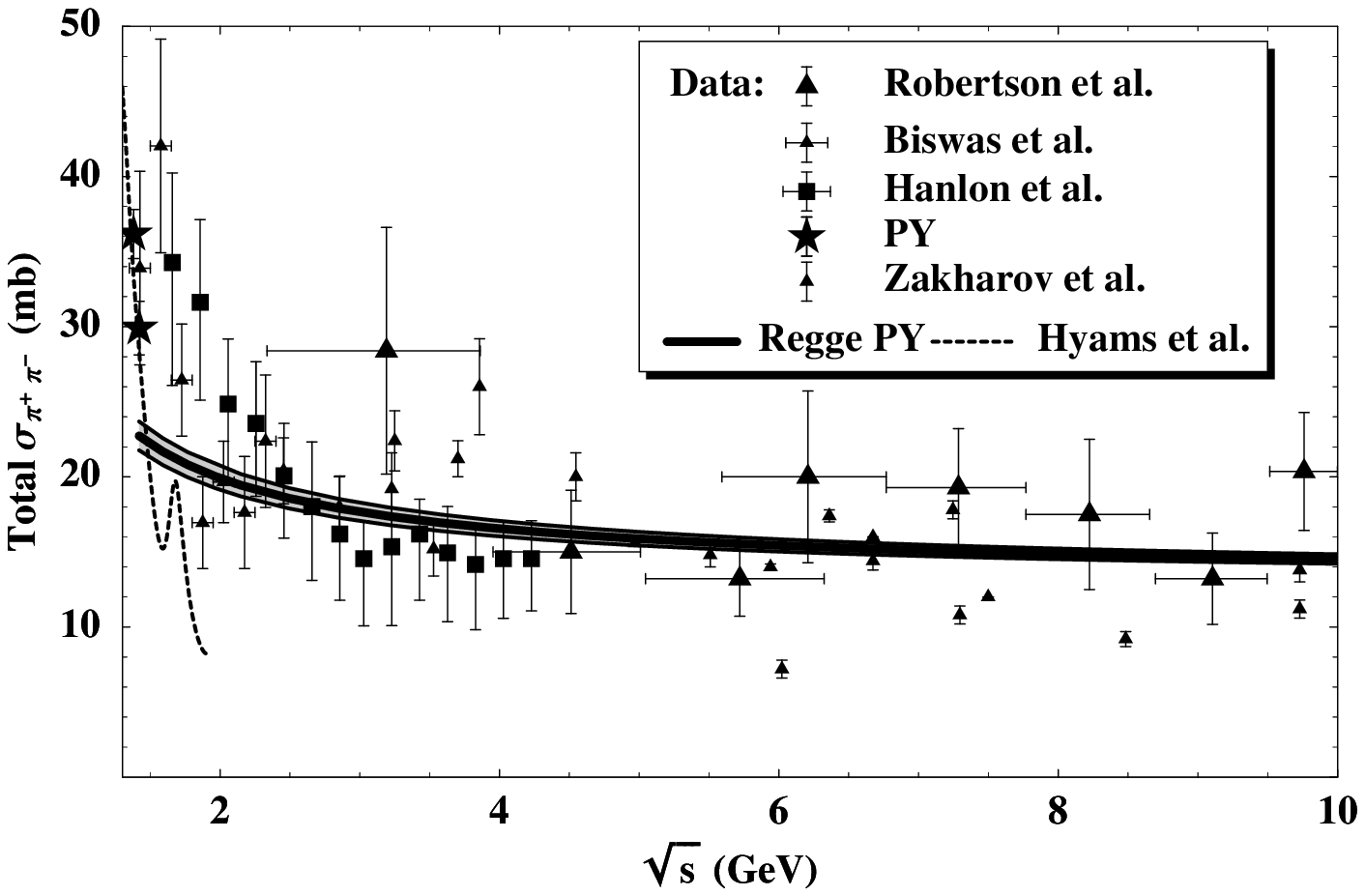}

\vspace*{-2.5cm}
\hspace*{-.5cm}
\begin{minipage}{1.0\linewidth}
{\footnotesize
{\sc \bf FIGURE 2} 
The $\pi\pi$ cross sections. Experimental points from \cite{14}.
The stars at 1.38 and 1.42 \gev\ (PY) are from the phase shift analysis 
of experimental data given in \cite{5}. 
Continuous lines, from 1.42 \gev\ (PY): Regge formula, with parameters as in 
\cite{4} (the gray bands cover the uncertainties).
Below 2 \gev, the dotted line 
corresponds to  the $\pi^+\pi^-$ cross section from 
 the Cern--Munich analysis of Hyams~et~al.\cite{10}
}
\end{minipage}

\vspace{.3cm}
\end{minipage}


\begin{minipage}{\textwidth}
\footnotesize
\begin{tabular}{|c|c|c|c|c|c|}
\hline
&$B_0$&$B_1$&$M_\mu$ (MeV)&${I_t=1}\atop{ \chi^2/\rm d.o.f.}$&
${\pi^0\pi^0}\atop{ \chi^2/d.o.f.}$
\\
\hline
PY,  Eq.(\ref{(2.3a)})&$21.04$&
$6.62$&$782\pm24$&$0.3$ &
$3.5$
\\\hline
$K\; {\rm decay\;  only}$
&$18.5\pm1.7$&$\equiv0$&$766\pm95$
&$0.2$  &$1.8$
\\ \hline
K decay + Grayer B \cite{10}&
$22.7\pm1.6$ &$12.3\pm3.7 $&$858\pm15 $&
$1.0$&$2.7$
\\
\hline
K decay +Grayer C \cite{10}
&
$ 16.8\pm0.85$ & 
$-0.34\pm2.34$ & $787\pm9 $
&$0.4$&$1.0$
\\
\hline
K decay + Grayer E \cite{10}&
$21.5\pm3.6 $&$12.5\pm7.6 $&$1084\pm110 $&
$2.1$&$0.5$
\\
\hline
K decay + Kamisnki \cite{10}&
$ 27.5\pm3.0$&$21.5\pm7.4 $&$789\pm18 $&$0.3$&$5.0$
\\ \hline \hline
K decay + Grayer A \cite{10}&
$ 28.1\pm1.1$&$26.4\pm2.8 $&$866\pm6 $
&$2.0$&$7.9$
\\
\hline
K decay+ EM, s-channel \cite{10}&
$ 29.8\pm1.3$&$25.1\pm3.3 $&$811\pm7 $&$1.0$&$9.1$
\\
\hline
K decay+ EM, t-channel \cite{10}&
$ 29.3\pm1.4$&$26.9\pm3.4 $&$829\pm6 $&$1.2$&$10.1$
\\
\hline
K decay+Protopopescu VI \cite{9}&
$ 27.0\pm1.7$&$22.0\pm4.1 $&$855\pm10 $&$1.2$&$5.8$
\\
\hline
K decay+Protopopescu XII \cite{9}&
$ 25.5\pm1.7$&$18.5\pm4.1 $&$866\pm14 $&$1.2$&$6.3$
\\
\hline
K decay+ Protopopescu 3 \cite{9}&
$ 27.1\pm2.3$&$23.8\pm5.0 $&$913\pm18 $&
$1.8$&$4.2$
\\
\hline
\end{tabular}

\hspace*{-.5cm}
\begin{minipage}{1.0\linewidth}
\vspace*{.3cm}
  {\footnotesize {\sc \bf TABLE 1} Fits to K decays and 
different data sets of S0 phase shifts.
 PY is our global fit,    Eq.(\ref{(2.3a)}), and its   $B_0$ and
    $B_1$ uncorrelated uncertainties can be obtained from
    Eq.(\ref{(2.3a)}). Note that many of these sets have very large $\chi^2/d.o.f.$
for the $I_t=1$ and $\pi^0\pi^0$ dispersion relations below $950\mev$. 
Let us also remark that 
some of them (like that of Sol. E) have a relatively 
small  $\chi^2/d.o.f.$ just because it has huge uncertainties in its parameters.
Ideally one would require a low $\chi^2/d.o.f.$ with small uncertainties
in the parameters.}
  \label{tab:a1}
\end{minipage}
\end{minipage}

\vspace*{-.2cm}
\subsection{Checking  and improving amplitudes with forward dispersion relations}
\vspace*{-.3cm}

Let us study how well the previous fits to data
satisfy 
three independent scattering amplitudes. We choose the
$t$-symmetric or antisymmetric combinations,
that form a complete set: 
$F_{00}\equiv F(\pi^0\pi^0\to\pi^0\pi^0)$, 
$F_{0+}\equiv F(\pi^0\pi^+\to\pi^0\pi^+)$, and the 
$t$ channel isospin one amplitude, $F^{(I_t=1)}$.
Final uncertainties are small for the two first,
since they
depend  only on two isospin states, and 
their imaginary parts are sums of positive terms. 
Thus, we find two dispersion relations by choosing either $F=F_{00}$ or $F=F_{0+}$ 
 in 
\begin{eqnarray}
\real F(s)-F(4M_{\pi}^2)=
\dfrac{s(s-4M_{\pi}^2)}{\pi}\pepe\int_{4M_{\pi}^2}^\infty\dd s'\,
\dfrac{(2s'-4M^2_\pi)\imag F(s')}{s'(s'-s)(s'-4M_{\pi}^2)(s'+s-4M_{\pi}^2)}.
\label{(4.1a)}
\end{eqnarray}
By setting $s=2M^2_\pi$, and  $F=F_{00}$,  we find
a sum rule important to fix the Adler zeros.
\begin{equation}
F_{00}(4M_{\pi}^2)=F_{00}(2M_{\pi}^2)+
\dfrac{8M_{\pi}^4}{\pi}\int_{4M_{\pi}^2}^\infty\dd s\,
\dfrac{\imag F_{00}(s)}{s(s-2M_{\pi}^2)(s-4M_{\pi}^2)}.
\label{(4.1b)}
\end{equation}
Finally, for isospin unit exchange, which does not require subtractions, 
\begin{equation}
\real F^{(I_t=1)}(s,0)=\dfrac{2s-4M^2_\pi}{\pi}\,\pepe\int_{4M^2_\pi}^\infty\dd s'\,
\dfrac{\imag F^{(I_t=1)}(s',0)}{(s'-s)(s'+s-4M^2_\pi)},
\label{(4.3)}
\end{equation}
at threshold this is known as the Olsson sum rule.

Depending on the method we use to fit the S0 wave we find the results in Table~1,
where, we have separated on top those fits to data with a total
 $\chi^2/d.o.f.<6$ for the $\pi^0\pi^0$ and $I_t=1$ dispersion relations
up to 0.925 GeV, 
a fairly reasonable $\chi^2/d.o.f.$ since these
fits were obtained independently of the dispersive approach.

However, in Table 1 we also list the very frequently used 
$t$ and $s$-channel solutions of Estabrooks and Martin 
\cite{10}, those of Protopopescu {\it et al.}\cite{9}, 
from Table VI, VIII and table XII, as well as 
the solution A of Grayer {\it et al.} \cite{10}. 
Their  $I_t=1$ plus $\pi^0\pi^0$ dispersion relation
total $\chi^2/d.o.f.$ is surprisingly poor: 11.3, 10.1, 7, 6, 7.5, 9.9, respectively.
{\it Therefore, any result that relies heavily on these
sets should be taken very cautiously}.


We have also improved the previous low energy fits parameters
by fitting also the 
 dispersion relations up to 0.925~\gev, thus obtaining parameterizations
more compatible with analyticity and $s\,-\,u$ crossing. 
Improving from Eq.(\ref{(2.3a)}), we find, in $M_\pi$ units,
\begin{eqnarray}
{\rm S0};\; s^{1/2}\leq 2m_K:&&  B_0=17.4\pm0.5;\; B_1=4.3\pm1.4;
\cr && M_\mu=790\pm21\,\mev;\;
z_0=195\,\mev\;\hbox{[Fixed]};\cr
{\rm S2};\; s^{1/2}\leq 1.0:&& B_0=-80.8\pm1.7;\; B_1=-77\pm5;\;
z_2=147\,\mev\;\hbox{[Fixed]};\cr
{\rm P};\; s^{1/2}\leq 1.05:&& 
B_0=1.064\pm0.11;\; B_1=0.170\pm0.040;\;
M_\rho=773.6\pm0.9\;\mev;\cr
{\rm D2};\; s^{1/2}\leq 1.42:&&  
B_0=(2.9\pm0.2)\times10^3;\; B_1=(7.3\pm0.8)\times10^3;\;\cr
&&
B_2=(25.4\pm3.6)\times10^3;\;\Delta=212\pm19. \label{(4.4)}
\end{eqnarray}
The D0 and F waves 
do not change appreciably and we just refer to \cite{5}
for details.
In Fig.3 we show the improved curves for S0 and S2, 
and  that of D2 in Fig.1.c.

Concerning the improved fits to individual data sets, we get 
somewhat different results for S0, listed in Table 2 with
the $\chi^2/d.o.f.$ of each forward dispersion relation and the
standard deviations for the sum rule in Eq.(\ref{(4.1b)}) (which are more than four
for  K decay plus the Grayer B or E 
or Kaminski improved solutions). 
For other waves, no matter what S0 fit is used,
we find very similar values to those in Eq.(\ref{(4.4)}).
This can be checked in  Fig.3.b, where we show
the improved ``K decay + Grayer Sol. B'' S2 wave. Even though
it is the one for which we obtained the 
most different central values for the S0 wave compared with those
given in Eq.(\ref{(4.4)}), it falls within the uncertainty
of our improved solution. 

In summary, with forward dispersion relations, 
we have checked the consistency 
of different phase shift analyses
available in the literature. Surprisingly,
some of the most frequently used phase shift sets 
do not satisfy these dispersive constraints and sum rules, 
and should therefore be used cautiously.
We have provided a simple parameterization of $\pi\pi$ scattering
consistent  simultaneously 
with some data sets and all three forward dispersion relations, that
we hope could be of use for future
studies of $\pi\pi$ scattering.

\hspace*{-.3cm}
\begin{minipage}{1.0\linewidth}
\includegraphics[height=.335\textheight,angle=-90]{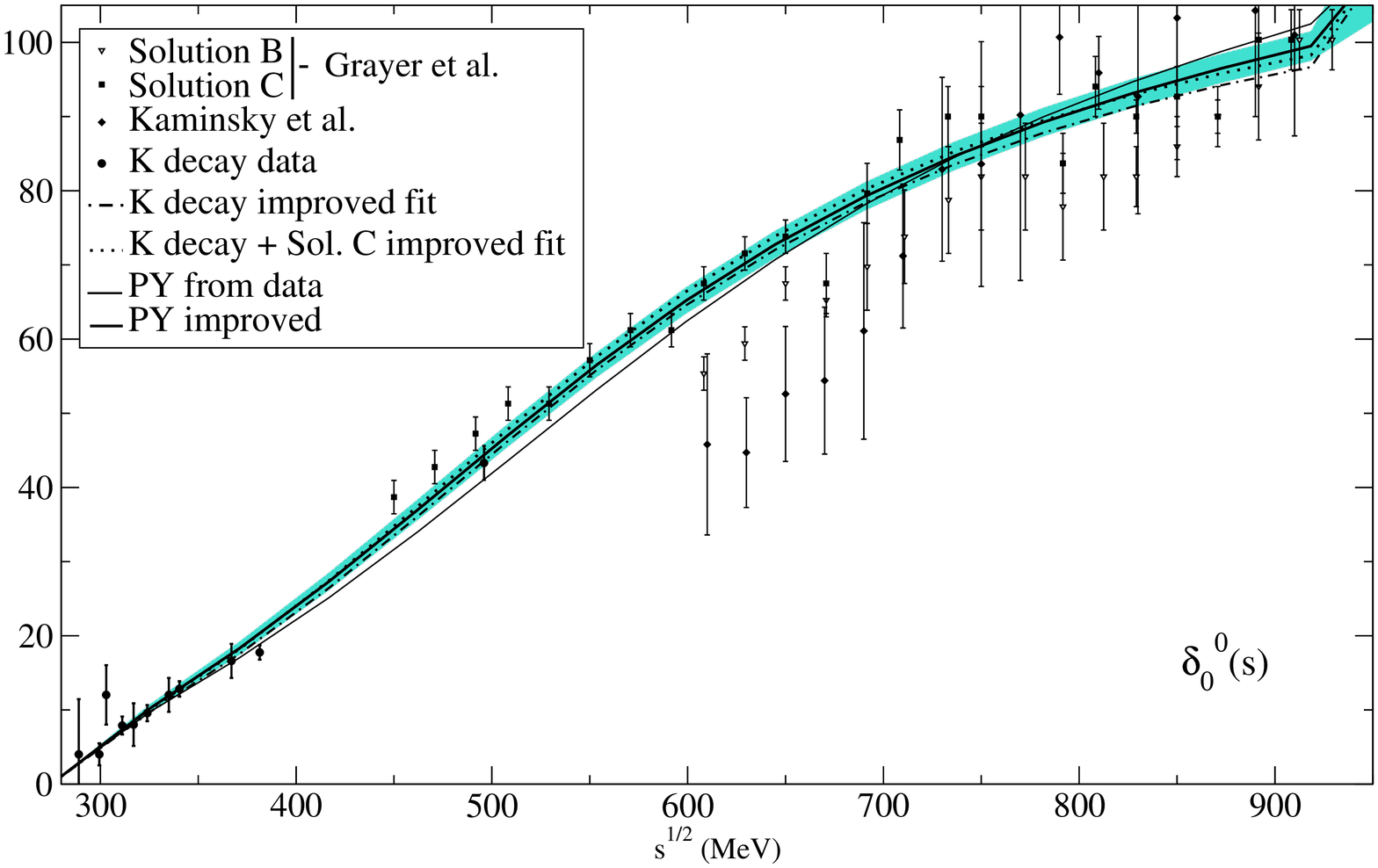}
\includegraphics[height=.335\textheight,angle=-90]{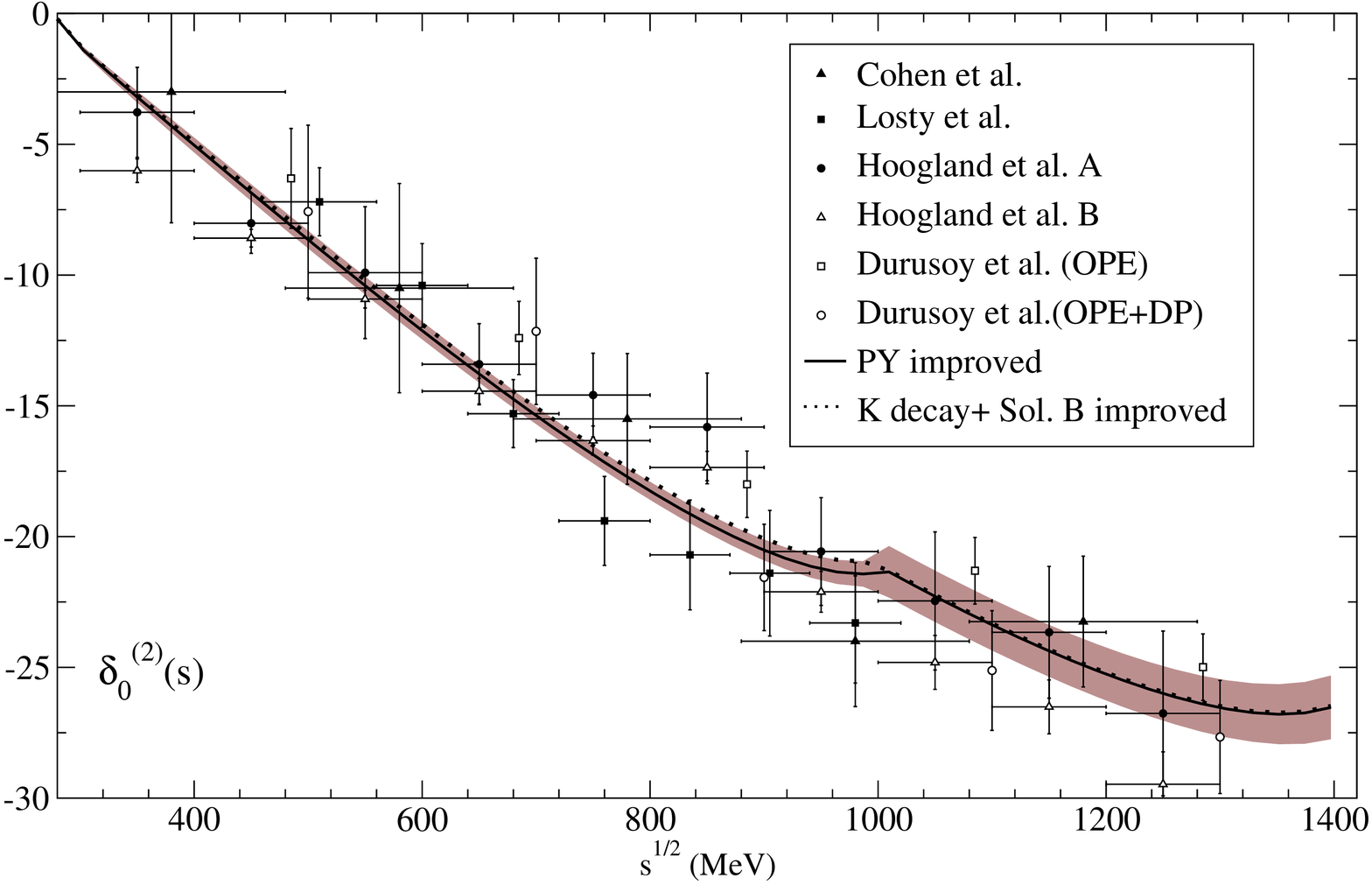}

{\footnotesize {\sc\bf FIGURE 3.} a) The improved S0 phase shift (PY improved, Eq.\ref{(4.4)}), 
the global fit (PY from data, the S0 in
Eq.~(\ref{(2.3a)})),
 and the 
{\sl improved} solutions ``$K$ decay only" and
 ``Grayer~C" of Table~2 (almost on top of PY improved). 
b) S2 improved Phase shift (PY improved,   
  Eq.~(\ref{(4.4)})); global fit (PY from data,  Eq.~(\ref{(2.3a)}))
and the improved
parameterization with K decays and So. B of Grayer et al.\cite{10}.}
\end{minipage}
\vspace{.2cm}

\hspace*{-.3cm}
\begin{minipage}{1.0\linewidth}
{\footnotesize {\sc\bf TABLE 2.}
Improved fits (only those already with $\chi^2/dof<6$ in Table~1).
Although errors are given for the Adler zero, we fix it 
when evaluating other errors, to break the otherwise very large 
correlations.}
\footnotesize
\begin{tabular}{|c|c|c|c|c|c|c|}
\hline
Improved& Improved  &$K$ decay only& $K$ decay+ &
$K$ decay+ &$K$ decay+ &$K$ decay+ \\
fits:&PY, Eq.\ref{(4.4)}&&Grayer C& Grayer B&Grayer E&Kami\'nski\\
\hline
$B_0$ & $ 17.4\pm0.5$&$16.4\pm0.9$&$16.2\pm0.7$ &$ 20.7\pm1.0$
&$ 20.2\pm2.2$&$ 20.8\pm1.4$\\
$B_1$&$4.3\pm1.4$&$\equiv0$&$0.5\pm1.8 $&$11.6\pm2.6 $&$8.4\pm5.2 $
&$13.6\pm43.7 $\\
$M_\mu\,$(MeV)&$790\pm30$&$809\pm53$&$788\pm9 $
&$861\pm14 $&$982\pm95 $&$798\pm17$ \\
$z_0\,$(MeV)&$195\pm30$&$182\pm34$&$182\pm39 $&$233\pm30 $
&$272\pm50 $&$245\pm39 $\\
\hline
${\displaystyle I_t=1,\; \chi^2/d.o.f.}$&0.40&0.30&0.37&0.37&0.60&0.43\\
\hline
${\displaystyle \pi^0\pi^0,\; \chi^2/d.o.f.}$&0.66
&0.29&0.32&0.83&0.09&1.08\\
\hline
${\displaystyle \pi^+\pi^-,\, \chi^2/d.o.f.}$&
1.62&1.77&1.74&1.60&1.40&1.36\\
\hline
Eq.(\ref{(4.1b)})&1.6$\sigma$&1.5$\sigma$&1.5$\sigma$&4.0$\sigma$&6.0$\sigma$&
4.5$\sigma$\\
\hline
\end{tabular}
\end{minipage}

\end{document}